\documentclass[useAMS,usenatbib]{mn2e}
\usepackage{amsmath}
\usepackage[pdftex]{graphicx}

\newcommand{\ba}{\mbox{\boldmath{$\alpha$}}}
\newcommand{\bt}{\mbox{\boldmath{$\theta$}}}
\newcommand{\bb}{\mbox{\boldmath{$\beta$}}}
\newcommand{\f}{{\cal F}}
\newcommand{\g}{{\cal G}}
\newcommand{\bx}{{\mathbf x}}
\newcommand{\by}{{\mathbf y}}

\begin{document}

\label{firstpage}
\title[An Aperture Mass Statistic for Flexion]{Detecting Mass Substructure in Galaxy Clusters: An Aperture Mass Statistic for Gravitational Flexion} \author[A. Leonard, L. J. King and S.M. Wilkins] {Adrienne
  Leonard\thanks{Email: leonard@ast.cam.ac.uk}, Lindsay
  J. King, and Stephen M. Wilkins\\Institute of Astronomy, University of Cambridge, Cambridge, CB3 0HA}

\date{}
\maketitle

\pagerange{\pageref{firstpage}--\pageref{lastpage}} \pubyear{2009}

\begin{abstract} Gravitational flexion has been introduced as
  a technique by which one can map out and study substructure in
  clusters of galaxies. Previous analyses involving flexion have
  measured the individual galaxy-galaxy flexion signal, or used either
  parametric techniques or a KSB-type inversion to reconstruct the
  mass distribution in Abell 1689. In this paper, we present an
  aperture mass statistic for flexion, and apply it to the lensed
  images of background galaxies obtained by ray-tracing simulations
  through a simple analytic mass distribution and through a galaxy
  cluster from the Millennium simulation.  We show that this method is
  effective at detecting and accurately tracing structure within
  clusters of galaxies on sub-arcminute scales with high
  signal-to-noise even using a moderate background source number
  density and image resolution. In addition, the method provides much more
  information about both the overall shape and the small-scale
  structure of a cluster of galaxies than can be achieved through a weak lensing 
  mass reconstruction 
  using gravitational shear data. Lastly, we discuss how the
  zero-points of the aperture mass might be used to infer the masses
  of structures identified using this method.
\end{abstract}


\section{Introduction}
Galaxy clusters are the most massive bound structures in the universe,
some in excess of $10^{15}M_{\odot}$, with about 90\% of their mass in
the form of dark matter. The mass function of clusters provides a
sensitive test of our cosmological model (e.g. Bahcall \& Fan 1998;
Eke et al. 1998), and the mass function of substructure provides
important insight into galaxy formation in dense environments (e.g. De
Lucia et al. 2004; Taylor \& Babul 2005).

In a universe dominated by Cold Dark Matter (CDM) such as our
concordance cosmology, $\Lambda$CDM, hierarchical structure formation
takes place with the growth of collapsed objects progressing via
merging of smaller objects. In non-hierarchical structure formation,
taking place in a universe dominated by Hot Dark Matter (HDM) for
example, the first haloes form via monolithic collapse. Recent N-body
simulations of halo formation in a HDM dominated universe by Wang \&
White (2008) show that it is the mass {\em substructure} content of
haloes - rather than most other halo properties such as shape or spin
parameter - that is markedly different from that seen in CDM.

Gravitational lensing, being insensitive to whether matter is luminous
or dark and to its dynamical state, is an ideal probe of
substructure. Natarajan, De Lucia \& Springel (2007) have obtained
constraints on the substructure in five massive galaxy clusters using
weak and strong gravitational lensing observations, finding comparable
levels ($\sim 10-20\%$) to that seen in high-resolution simulated
clusters.

In addition, Leonard et al. (2007) and Okura et al. (2008) have used
flexion measurements to detect substructure in Abell 1689. While the
Okura et al. reconstruction uses a nonparametric, KSB-type
(Kaiser, Squires \& Broadhurst, 1995) reconstruction technique, the
Leonard et al. reconstruction relies on a parametric modelling of
known cluster members. Each of these methods has its disadvantages; using the finite field inversion technique of Okura et al., it is difficult to accurately characterise the noise properties of the mass maps produced, and thus to assess the significance of the
detections, whilst the Leonard et al. reconstruction requires detailed knowledge of the locations of the structures responsible for the lensing. Since flexion studies in galaxy clusters are highly
sensitive to the substructure within the cluster, these studies cannot
be compared to weak or strong lensing reconstructions, or to mass
models of the cluster as a whole, to assess their performance.

In this paper, we derive an aperture mass statistic for gravitational
flexion in direct analogy to that for gravitational shear (e.g. Schneider 1996;
Schneider et al. 1998), to enable us to use measurements of flexion to
determine the locations and statistical significance of mass peaks in
clusters of galaxies without the need for any parametric modelling or
comparison with complementary lensing studies. We begin with a review
of the flexion formalism, outline how flexion is measured and
techniques that have previously been used to map the mass distribution
in Abell 1689. In Section 3 the aperture statistic for flexion is
derived, and in Section 4 we describe the ray-tracing simulations that
have been carried out and how the aperture statistic is calculated
from the synthetic data. Section 5 presents results on the performance
of the statistic, and in Section 6 we discuss our findings and
conclude.

Throughout this paper we use a matter density parameter $\Omega_{\rm
  m}=0.27$, dark energy density parameter $\Omega_{\Lambda}=0.73$
(with equation of state parameter $w=-1$) and Hubble parameter
$H_{0}=71$km$\,$s$^{-1}\,$Mpc$^{-1}$.
\section{Review of Flexion Formalism}

In traditional weak lensing studies, the lens equation is approximated
as linear, and lensed images exhibit a purely elliptical distortion
aligned tangential to the lens. Linearisation of the lens equation
assumes that there is no variation of the lens field over the scale of
the lensed image. If this field is allowed to vary smoothly, the lens
equation becomes non-linear (see Goldberg \& Bacon 2005):
\begin{equation}
\beta_i \simeq A_{ij}\theta_j+\frac{1}{2}D_{ijk}\theta_j \theta_k,
\end{equation}
where $\beta$ is the coordinate in the source plane, $\theta$ is the
lensed coordinate, ${\bf A}$ is the magnification matrix
\begin{equation}
{\bf A}= \left(
\begin{array}{clrr}
        1-\kappa-\gamma_1 & -\gamma_2\\ -\gamma_2 & 1-\kappa+\gamma_1
\end{array}
\right)\,,
\end{equation}
$\kappa$ is the convergence, $\gamma_{1}, \gamma_{2}$ are the
components of the (complex) gravitational shear (a polar quantity),
and $D_{ijk}=\partial_kA_{ij}$. The ${\bf D}$ operators can be related
to two measurable quantities, first and second flexion, by
\begin{equation}
  D_{ij1}=-\frac{1}{2}\left( 
    \begin{array}{clrr} 
      3\f_1+\g_1 & \f_2+\g_2 \\
      \f_2+\g_2 & \f_1-\g_1
    \end{array}
  \right),\nonumber
\end{equation}
\begin{equation}
  D_{ij2}=-\frac{1}{2}\left(
    \begin{array}{clrr}
      \f_2+\g_2 & \f_1-\g_1 \\ \f_1-\g_1 & 3\f_2-\g_2
    \end{array}
  \right),
\end{equation}
where $\f=\f_1+i\f_2=\partial\kappa$, $\g=\g_1+i\g_2=\partial\gamma$,
and $\partial=\partial_1+i\partial_2$ is the differential operator
defined in Bacon et al. (2006). 

First flexion, $\f$, transforms as a vector, directly probing the
gradient of the convergence, and gives rise to a skewness in the light
distribution of the lensed image aligned radially with respect to the
lens. Second flexion, $\g$ has $m=3$ rotational symmetry and gives
rise to an arciness in the lensed image, aligned tangential to the
lens. Goldberg \& Leonard (2007) and Leonard et al. (2007) found that
second flexion is significantly more difficult to measure than first
flexion, thus for the
purposes of this paper, we will consider only first flexion.

There are two distinct methods by which flexion can be measured. The
first, described in detail in Goldberg \& Bacon (2005), Bacon et
al. (2006) and Massey et al. (2007) involves the decomposition of
lensed images into an orthogonal basis set, shapelets (see
e.g. Bernstein \& Jarvis 2002, Refregier 2003, Refregier \& Bacon 2003, Massey \& Refregier
2005), which are related to two-dimensional reduced Hermite or
Laguerre polynomials. One advantage of this method is that the various
lensing operators can be expressed rather simply in terms of the
quantum-mechanical raising and lowering operators, and produce rather
compact transfers of power between shapelet modes. Another advantage
is that an explicit deconvolution of the telescope point spread
function (PSF) can be carried out rather straightforwardly in this
method.

An alternative technique has been formulated by Okura et al. (2007)
and subsequently refined and extended in Goldberg \& Leonard (2007)
and Okura et al. (2008). This technique involves measuring higher
order moments of the brightness distribution of the lensed image
(HOLICs), and using these moments to estimate the flexion. While less
compact on paper, the HOLICs method has the advantage that the
computational analysis time is dramatically shorter than that involved with
shapelets, particularly on large, well-resolved lensed
images. In addition, incorporating a weighting function in the
measurement of the HOLICs reduces the impact of noise in the image, as
well as light contamination from nearby sources, and the latest work
by Okura et al. (2008) describes a method whereby the effects of both
an isotropic and an anisotropic PSF can be corrected for. 

The shapelets technique has been used successfully to measure
galaxy-galaxy flexion in a sample of field galaxies in the Deep Lens
Survey (Goldberg and Bacon 2005), while two different implementations
of the HOLICs formalism have been used to measure flexion in images of
Abell 1689 (Leonard et al. 2007, Okura et al. 2008), where it has been
shown that flexion is particularly sensitive to substructure in
clusters of galaxies. While the Leonard et al. reconstruction includes
a description of the noise properties of their reconstructed
convergence map, the technique relies on accurate knowledge of the
locations of the cluster members responsible for the measured flexion
signal, and takes no account of the smooth component of the cluster
potential, nor does it account for the possible presence of dark
haloes within the cluster. The Okura et al. reconstruction has the
advantage of using a nonparametric technique, and thus not requiring
any a priori assumptions about the cluster. However, it is very
difficult with this technique to accurately describe the noise
properties of the convergence map generated.

It is in this context that an aperture mass statistic, which includes
a straightforward description of the noise properties of the mass
peaks detected, is important.

\section{The Aperture Mass Statistic}

To derive the aperture mass statistic for flexion, we closely follow
the work of Schneider (1996), in which the formalism for the
generalised aperture mass statistic for shear was first laid out. The
aperture mass is defined as:
\begin{equation}
\label{eq:apmass}
m(\bx_0)= \int d^2\bx\ \kappa(\bx+\bx_0)\ w(|\bx|).
\end{equation}
We aim to express this statistic in terms of some measurable quantity,
namely the measured flexion. 

The convergence is related to the first flexion by (see Appendix \ref{ap:kernels})
\begin{equation}
\label{eq:kfl}
  \kappa(\bx)=\frac{1}{2\pi}\Re\left[\int d^2\bx^\prime\
    E_\f^\ast(\bx-\bx^\prime)\f(\bx^\prime)\right]+\kappa_0, 
\end{equation}
where
\begin{equation}
E_\f=\frac{1}{X^\ast},
\end{equation}
and $X=x_1+ix_2$. Thus, we can redefine the aperture mass as
\begin{eqnarray}
  \lefteqn{m(\bx_0)=}\nonumber\\
  & &\frac{1}{2\pi}\left(\Re \left[\int d^2\bx\ w(x)\int
      d^2\bx^\prime\ E_\f^\ast
      (\bx-\bx^\prime+\bx_0)\f(\bx^\prime)\right] \right.\nonumber\\
  & & \left.+\int d^2\bx\ w(x)\kappa_0\right).
\end{eqnarray}
Schneider (1996) notes that the aperture mass can be made
independent of the constant quantity $\kappa_0$ if we require that the
mass filter function $w(x)$ is compensated, i.e.
\begin{equation}
\label{eq:comp}
\int_0^\infty x\ w(x) dx=0.
\end{equation}

Under this condition, the aperture mass for flexion can be
redefined in terms of the ``E-mode'' (radially aligned) flexion as
follows (see Appendix \ref{ap:filt} for the complete derivation):
\begin{equation}
\label{eq:mflex}
m(\bx_0)=\int d^2\by\ \f_E(\by;\bx_0)\ Q_\f(y),
\end{equation}
where
\begin{equation}
\label{eq:qwflex}
Q_\f(y)=-\frac{1}{y}\int_0^y\ x\ w(x)\ dx.
\end{equation}

In direct analogy with the shear aperture mass statistic, the expected
signal to noise ratio, ${\cal S}$, achievable for a given flexion filter
function can be calculated by taking an ensemble average over the probability distribution for the background galaxy positions, which gives
\begin{equation}
  \label{eq:snflex} {\cal S}=\frac{2\sqrt{\pi
      n}}{\sigma_\f}\frac{\int_0^R\ x\ Q_\f(x)\left\langle
      \f_E\right\rangle(x)\ dx} {\sqrt{\int_0^R\ 
      x\ Q^2_\f(x) dx}}, 
\end{equation}
where $n$ is the number density of background sources within the
aperture, and $\sigma_\f$ is the dispersion in flexion
measurements. Goldberg \& Leonard (2007) found $\sigma_{a|\f|}=0.03$, implying $\sigma_\f\sim 0.28/\arcsec$. We note that this is likely to be an overestimate of the true, unlensed dispersion in flexion values, however, as this estimate is based on the flexion measured in galaxies within a moderate lensing field.
\subsection{Choosing a Filter}
We consider two different families of filter functions for the flexion
aperture mass statistic. These filters are able to attain a very
similar peak signal-to-noise; however, the dependence of the signal to
noise ratio on various factors (such as the scale of the lens
substructures and the size of the aperture being used) differs
significantly between the two sets of filters, thus their respective domains of applicability differ.

It is important to note that the signal to noise attainable for a given lens profile is maximised when the flexion filter function traces the expected flexion signal. Thus the optimal choice of filter function is strongly dependent on the flexion profiles of the structures being studied. The filter functions presented in this paper have not been chosen to be optimal for a given lens profile; rather they are designed to have rather broad applicability, thus allowing us to detect structures without requiring a priori knowledge of the profiles of these structures.
\subsubsection{Piecewise-continuous Filters}
\label{subsubsec:96}
Schneider (1996) describes a generalised piecewise-continuous,
compensated mass filter function, from which he derives a shear filter function:
\begin{equation}
w_\gamma(x)=
\begin{cases}
 1 & {\scriptstyle x<\nu_1R}\\ \\
    \frac{1}{1-c}\left(\frac{\nu_1R}{\sqrt{(x-\nu_1R)^2+(\nu_1R)^2}}-c\right)
    & {\scriptstyle \nu_1R<x<\nu_2R}\\ \\
    \frac{b}{R^3}(R-x)^2(x-\alpha R) &
    {\scriptstyle \nu_2R<x<R}
    \end{cases},
\end{equation}
where $\nu_1 R$ is an inner radius, usually taken to be small compared to the size of the aperture, and $\nu_2 R$ is an outer radius, usually taken to be close to the aperture scale $R$. The constants $\alpha$, $b$ and $c$ are calculated using the
constraints that $w_\gamma(x)$ and $\partial w_\gamma(x)/\partial x$ must be continuous at
$x=\nu_1R$ and $x=\nu_2R$, and that $w_\gamma(x)$ must be compensated. 

Noting that the mass filter function for flexion is roughly proportional to the derivative of the mass filter function for shear, we choose
\begin{equation}
w_\f(x)=
\begin{cases} 1 & {\scriptstyle x<\nu_1R}\\ \\ \frac{1}{1-c}
  \left(\frac{(\nu_1R)^3} {\left((x-\nu_1R)^2+(\nu_1R)^2\right)^{\frac{3}{2}}}-c
  \right) & {\scriptstyle \nu_1R<x<\nu_2R} \\ \\
  \frac{b}{R^3}(R-x)^2(x-\alpha R) 
  & \scriptstyle{\nu_2R<x<R}, \end{cases},
\end{equation}
where, again, the constants can be computed by requiring that $w(x)$
and $\partial w(x)/\partial x$ be continuous at $x=\nu_1R$ and
$x=\nu_2R$, and that $w(x)$ be compensated.
Applying equation \ref{eq:qwflex}, we find that
\begin{eqnarray}
  Q_\f(x\le \nu_1R)=-\frac{x}{2},\nonumber
\end{eqnarray}
\begin{eqnarray}
 \lefteqn{Q_\f(\nu_1R\le x\le\nu_2R)=}\nonumber\\
  & &-\frac{(\nu_1R)^2\left(3+\frac{2(x-2\nu_1R)}{\sqrt{(x-\nu_1R)^2+
          (\nu_1R)^2}}\right)-cx^2}{2(1-c)x},\nonumber
\end{eqnarray}
\begin{eqnarray}
\label{eq:q_96}
\lefteqn{Q_\f(x\ge\nu_2R)=
  -\frac{R^2\left(\nu_1^2\left(3+\frac{2(\nu_2-2\nu_1)}
        {\sqrt{(\nu_2-\nu_1)^2+\nu_1^2}}\right)-c\nu_2^2\right)}{2(1-c)x}}
\nonumber\\ 
\nonumber\\ 
&-\frac{b}{20R^3x}\left( 4(x^5-(\nu_2R)^5)
  -5(2+\alpha)R(x^4-(\nu_2R)^4)\right)\nonumber\\ \nonumber\\
&-\frac{b}{6Rx}\left(2(1+2\alpha)(x^3-(\nu_2R)^3) -3\alpha
  R(x^2-(\nu_2R)^2)\right).\nonumber\\
\end{eqnarray}

\begin{figure}
\includegraphics[width=0.22\textwidth]{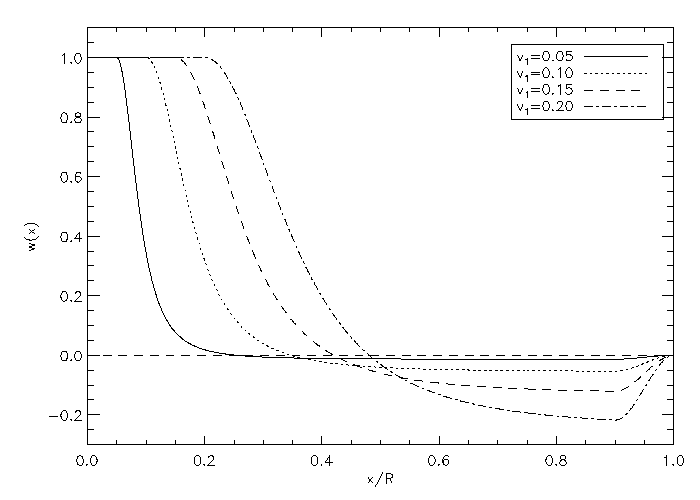}
\includegraphics[width=0.22\textwidth]{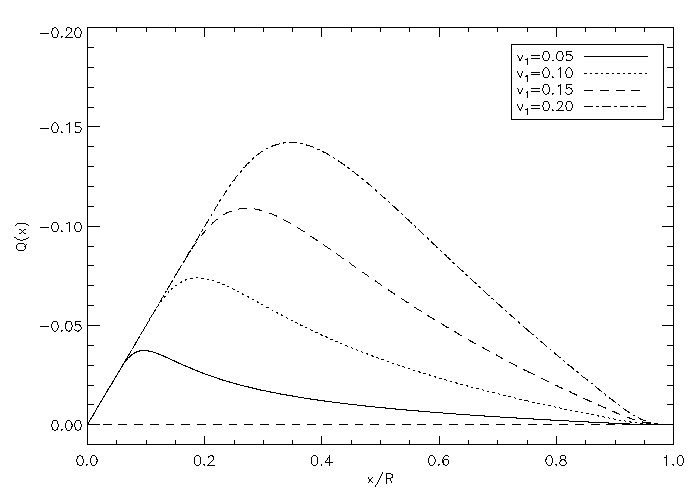}
\caption{\textit{Left Panel:} The piecewise-continuous mass filter function
  described in \S~\ref{subsubsec:96} plotted as a function of $x/R$,
  where R is the aperture radius, for $\nu_2=0.95$. \textit{Right Panel:} The
  corresponding flexion filter functions.\label{filters1996}}
\end{figure}

Fig.~\ref{filters1996} shows the mass and flexion filter functions derived
above for various choices of $\nu_1$, taking $\nu_2=0.95$. The flexion filter drops of roughly as $x^{-1}$ for $\nu_1 R\le x\le \nu_2R$, thus is not optimised for either an isothermal profile or an NFW profile (and, in fact, would only be optimal for a profile with $\kappa(x) \propto \log(x)$). However, as it has a rather wide distribution, we expect to be able to detect structures with a broad range of profiles using this filter. 

\subsubsection{Polynomial Filters}
\label{subsubsec:98}

A more compact family of (continuous) mass filter functions was described
by Schneider et al. (1998) as:
\begin{equation}
\label{eq:w_98}
w(x)=A\frac{(2+l)^2}{\pi}\left(1-\frac{x^2}{R^2}\right)^l\left(\frac{1}{2+l}-
  \frac{x^2}{R^2}\right),
\end{equation}
where $A$ is a factor arising from the normalisation of the associated
flexion filter function. This family of functions drops off with order $l$ as
$x\rightarrow R$, and were designed to be applicable for cosmic shear studies. Thus, the filter functions are not optimised, but should be sensitive to a range of mass profiles. 

The associated flexion filter functions are given by
\begin{equation}
\label{eq:q_98}
Q_\f(x)=
-A\frac{2+l}{2\pi}x\left(1-\frac{x^2}{R^2}\right)^{1+l}.
\end{equation}
It is clear from dimensionality considerations that $Q_\f$ must
have dimensions of length, as the product $\f Q_\f$ must be
dimensionless and flexion itself carries dimensions of inverse
length. Thus, we choose the normalisation constant, $A$, such that
\begin{equation}
\label{eq:leonardnorm}
\frac{2\pi}{R^3}\int_0^R x\ Q_\f(x) dx = 1.
\end{equation}
Thus,
\begin{equation}
\label{eq:normconst}
A=\frac{4}{\sqrt{\pi}}\frac{\Gamma\left(\frac{7}{2}+l\right)}{\Gamma(3+l)}.
\end{equation}

\begin{figure}
\includegraphics[width=0.22\textwidth]{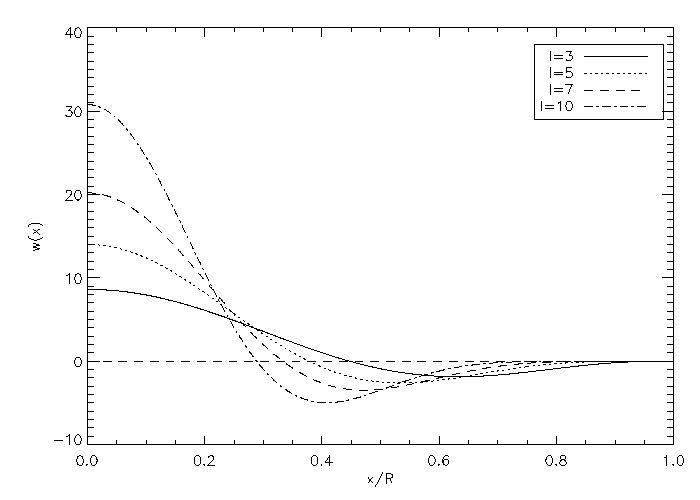}
\includegraphics[width=0.22\textwidth]{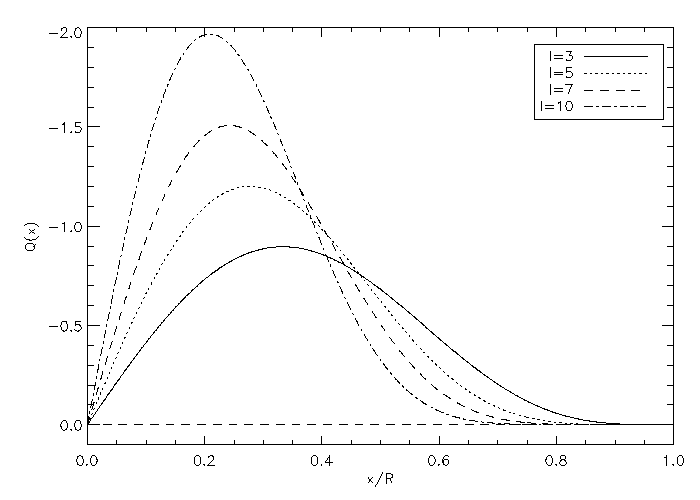}
\caption{\textit{Left Panel:} The polynomial mass filter function described
  in \S~\ref{subsubsec:98} plotted as a function of $x/R$. \textit{Right Panel:} The corresponding flexion
  filter functions.\label{filters1998}}
\end{figure}

These mass and flexion filter functions are plotted in Fig.~\ref{filters1998} for various values of $l$. 

\subsection{Expected Signal to Noise}
\label{subsec:sn}

The expected signal to noise achievable for a given filter function can be calculated according to equation \ref{eq:snflex}. This turns out to be quite a straightforward calculation using the polynomial flexion filter functions (equation \ref{eq:q_98}) and under the assumption of a singular isothermal sphere lens profile ($\f_E(x)=\theta_E/2x^2$), where we find the signal to noise to be given by:
\begin{equation}
\label{eq:snth}
{\cal S}=\frac{\theta_E}{R}\frac{\sqrt{n}}{\sigma_\f} \frac{\Gamma(2+l)}{\Gamma\left(\frac{5}{2}+l\right)}\sqrt{(2+l)(3+2l)} ,
\end{equation}
where $n$ is the background source density, $\sigma_\f$ is the intrinsic flexion dispersion, and $\theta_E$ is the Einstein radius of the lens, given by
\begin{equation}
\theta_E=4\pi\left(\frac{\sigma_v}{c}\right)^2\frac{D_{ls}}{D_s},
\end{equation}
where $D_s$ is the angular diameter distance to the source plane and $D_{ls}$ is the angular diameter  distance
between the lens and source planes. 

Thus, we can see that the expected signal to noise for this set of
filters decreases with increasing aperture size. This would
seem to imply that, in order to optimise the signal to noise, the
apertures should be made as small as possible. However, decreasing the aperture size
decreases the number of sources within each aperture. In a noisy image
where the flexion is not measured perfectly, increasing the number of
sources within an aperture will increase the statistical significance
of the measurement. In other words, having a small number of (noisy) flexion measurements within an aperture will artificially increase the value of $\sigma_\f$ within that aperture, thus decreasing the signal to noise achievable within the aperture. 

In addition, a small aperture size will inevitably result in many apertures over the field having no sources in them, particularly when using data with a low background source count. This will result in a patchy measured aperture mass with discontinuities seen where there are apertures containing no sources. It is thus necessary to use larger apertures than the signal to noise calculation might imply are optimal.

The above calculation also assumes that the flexion signal is measurable at $x=0$, and indeed more heavily weights the signal close to the centre of the aperture. Now, in practice, we will be unable to measure the flexion at very small distances from the centre of the lens because the lens itself will obscure any background objects directly behind it. In addition, the flexion profile described by equation \ref{eq:snth} becomes very large as $x\rightarrow 0$. A more reasonable model of the lens profile might be a softened isothermal sphere, in which the convergence is given by
\begin{equation}
\kappa(x)=\left(1+\frac{x^2}{2\theta_E^2}\right)\left(1+\frac{x^2}{\theta_E^2}\right)^{-\frac{3}{2}},
\label{eq:ssis}
\end{equation}
which gives rise to a flexion signal given by
\begin{equation}
\f_E(x)=-\frac{x}{\theta_E^2}\left(2+\frac{x^2}{2\theta_E^2}\right)\left(1+\frac{x^2}{\theta_E^2}\right)^{-\frac{5}{2}}.
\end{equation}

\begin{figure}
\center
\includegraphics[width=0.4\textwidth]{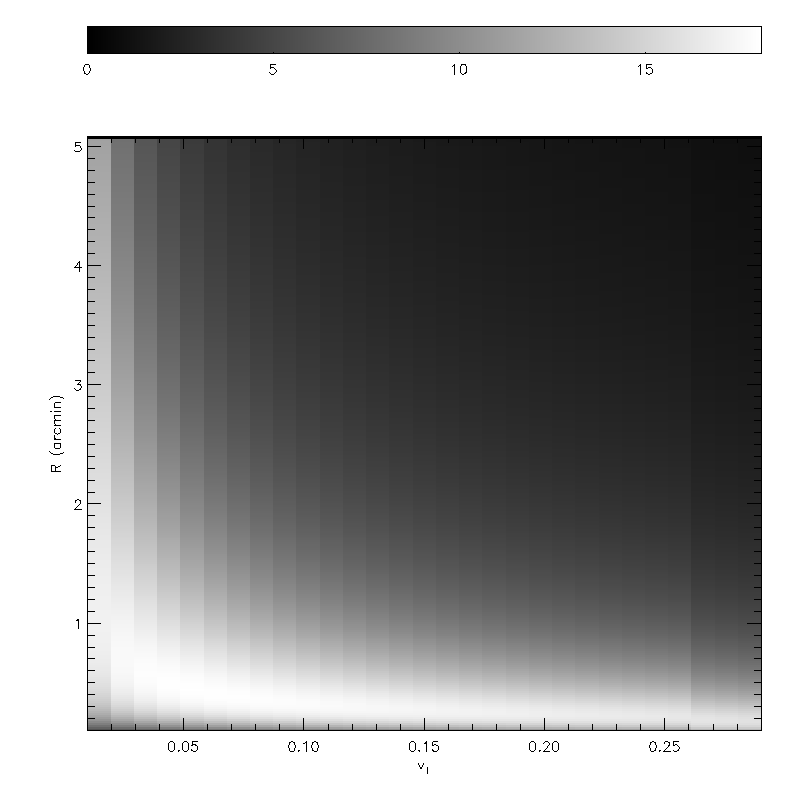}
\includegraphics[width=0.4\textwidth]{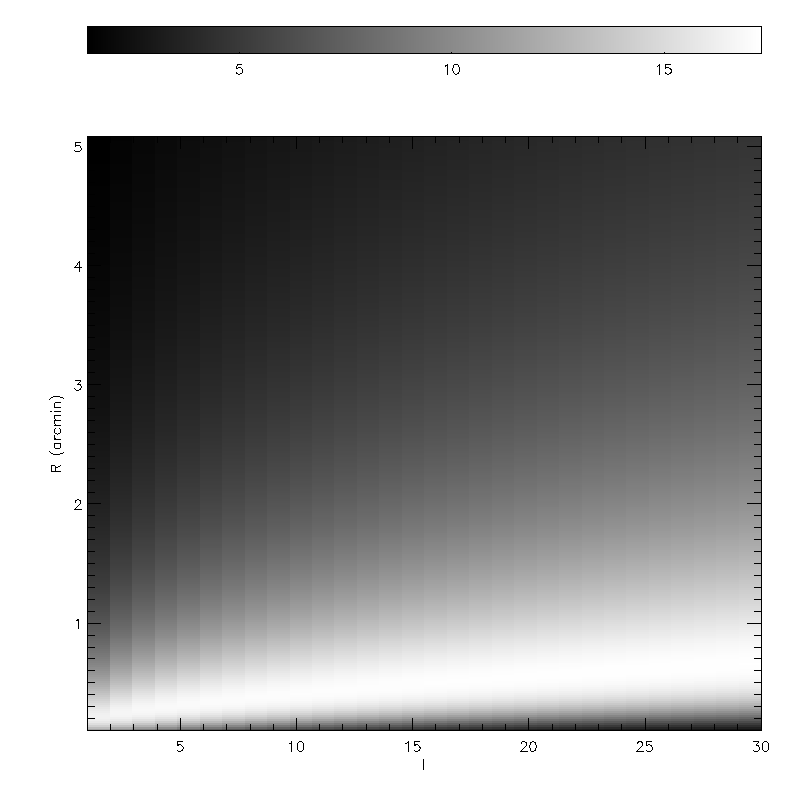}
\caption{\textit{Top Panel:} The expected signal to noise for the
  filter described in equation \ref{eq:q_96}, plotted as a function
  of $\nu_1$ and $R$. \textit{Bottom Panel: }The expected
  signal to noise for the family of filters described in equation
  \ref{eq:q_98} as a function of $l$ and $R$.
  \label{fg:sn}} 
\end{figure}

Figure \ref{fg:sn} shows the expected signal to noise for this softened isothermal profile
for the filter functions described by equations \ref{eq:q_96} and
\ref{eq:q_98}. For each, we assume a velocity dispersion of $\sigma_v=500$km s$^{-1}$, which gives rise to $\theta_E\simeq 5\arcsec.95$ assuming $D_{ls}/D_s=0.8$. We take $n=35$ arcmin$^{-2}$ and $\sigma_\f=0.1/\arcsec$. This is a low value compared to the dispersion in flexion measurements found by Goldberg and Bacon (2005) or Goldberg and Leonard (2007); however the simulated data with which we are concerned in this paper has a lower dispersion than that seen in real data, thus the value of $\sigma_\f$ taken here is representative of our data. In addition, the measured values found in Goldberg and Bacon (2005) and Goldberg and Leonard (2007) are likely to be overestimates of the ``true'' flexion dispersion, as these measurements were carried out in lens fields, thus we expect to see a flexion signal in these background galaxies. We find the peak attainable signal to noise for each filter to be $\sim18$, with a slightly higher peak signal to noise found using the piecewise-continuous flexion filter function. 

It is important to note that the expected signal to
noise ratio has been computed here under the assumption that there is
only one lens present in the regime being considered. However,
clusters of galaxies tend to be quite clumpy; thus we expect the
signal to noise achievable in clusters to be somewhat lower, though still appreciable.
 
It is clear from the figure that the signal to noise properties of these two filter functions differ significantly. Thus comparisons between the aperture mass measured using different filters provides a robust method for checking the accuracy of these maps. It is clear that for most values of $\nu_1$, the piecewise-continuous function favours smaller apertures; that is, apertures more closely matched with the characteristic size of the lens. The polynomial filters, on the other hand, favour larger apertures as the polynomial order is increased, and indeed should provide robust detections for a broad range of aperture sizes for any given choice of polynomial order $l$, particularly if $l$ is large. 

In addition, for a given choice of aperture size, increasing $l$ should decrease the minimum scale on which structures are resolved, as increasing $l$ decreases the radius at which the filter function reaches its peak. Similarly, decreasing the value of $\nu_1$ for the piecewise-continuous filter should decrease the minimum scale on which structures can be resolved. Thus it is possible with both these filters to resolve structures on both large and small scales by appropriately tuning the filter. This is important in clusters of galaxies, where one is interested in both the large-scale structure of the cluster potential, but also the smaller-scale substructures within it. 

\section{Application to Simulations}
We test our method on two different simulated lens systems using a
ray-tracing simulation to artificially lens background test galaxies
through a given convergence field.

\subsection{Ray-tracing Method}
\label{subsec:raytrace}
To carry out the ray-tracing, we first consider the mapping
between the source and lens planes. This is given by
$\bb=\bt-\ba(\bt)$, where $\bb$ are the coordinates in the source
plane, $\bt$ are the coordinates in the lens plane, and $\ba$ are the
deflection angles evaluated as a function of the lens plane
coordinates. In Fourier space, the convergence and the deflection
angle are related by
\begin{equation}
\label{eq:fft}
\tilde{\alpha_i}=-\frac{2ik_i}{k_1^2+k_2^2}\tilde{\kappa}.
\end{equation}

Thus, in order to calculate the deflection angles arising due to a given lens convergence, one simply needs to Fourier transform the convergence field, apply equation \ref{eq:fft}, and then compute the inverse Fourier transform. In order to achieve sufficient numerical accuracy and avoid edge effects, zero-padding is applied surrounding the convergence field. For all the simulations described in this section, we use a padding on all four sides of the image that is 3.5$\times$ the size of the input image.

\subsection{Analytic Simulation}
\label{subsec:analytic}
As a first test of the aperture mass technique, a purely analytic lens convergence is generated by laying down a number of randomly
distributed, circular test ``galaxies" with a convergence defined by equation \ref{eq:ssis}, and with a pixel scale of $0\arcsec .5$ pix$^{-1}$. For these simulations, we take $D_{ls}/D_s=0.8$, thus for a source redshift of $z_s=1.0$, the lens will be at a redshift of $z_l\simeq 0.16$. 

We test our method on both a single lens galaxy and a group of galaxies. The velocity dispersion for the single lens case is taken to be $\sigma_v=500$km s$^{-1}$, while for the group simulation, the velocity dispersion of each lens galaxy is drawn randomly from a Gaussian
distribution with a mean of 220\,km\,s$^{-1}$ and a standard deviation of
20\,km\,s$^{-1}$. The pixel locations of each test galaxy in the group are also
drawn from a Gaussian distribution centred on the centre of the image
grid, and with a standard deviation set to $N_{\rm pix}/10$, where $N_{\rm pix}$ is the width of the image in pixels. 

\subsection{N-body Simulation}
\label{subsec:nbody}
The advantage of using a purely analytic potential is that the method
can be tested and checked for accuracy at various stages of the analysis
pipeline, as the expected flexion and deflection angles can be calculated
analytically. However, this method does not generate a particularly
realistic model cluster of galaxies. 

For this purpose, we make use of
a cluster extracted from the Millennium simulation (Springel et al. 2005) at a redshift of $z_l=0.21$. The resolution of the cluster image used is 5$h^{-1}$ kpc/pixel ($2\arcsec .07$ pix$^{-1}$), corresponding to the gravitational softening of the simulation, and the projections are computed using a nearest grid point interpolation. The cluster has a characteristic radius $R_{200}=2.0\ h^{-1}$Mpc and a mass of $M_{200}=1.23\times 10^{15}\ h^{-1}M_\odot$. We extract the central 1.0$h^{-1}$Mpc of this cluster for use with our ray-tracing simulation. 

It is important to note that, when carrying out the ray-tracing using this cluster as our lens, we do not apply any smoothing or inter-pixel interpolation prior to computing the deflection angles across the lens plane grid. As the cluster data is quite noisy, and the pixel scale is significantly larger than that used for the simulated background galaxies, we expect this inter-pixel noise to result in noisier flexion data than that found when using a lens simulated as described in \S~\ref{subsec:analytic}.

\subsection{Background Galaxy Population}
\label{subsec:bg}
For all the ray-tracing simulations, we use an image resolution of $0.1\arcsec $pix$^{-1}$, and a background number density of 35 sources/arcmin$^2$. This may seem to be a rather optimistic value compared to typical ground based observations, and indeed when compared to the analysis of Abell 1689 carried out by Okura et al. (2008) using Subaru data, in which they used 5 sources/arcmin$^2$. However, the Leonard et al. (2007) analysis used $75$/arcmin$^2$, which is rather high, but typical of space-based observations. Further, current and upcoming surveys and instruments will regularly be able to achieve a background source count of in excess of $35$/arcmin$^2$. Some examples are the Large Synoptic Survey Telescope (LSST\footnote{LSST Dark Energy Task Force Whitepaper: http://www.lsst.org/Science/docs/LSST\_DETF\_Whitepaper.pdf}), which should offer 50 sources/arcmin$^2$, the Canada-France-Hawaii Telescope (CFHT) Legacy Survey, which obtains 35 sources/arcmin$^2$ (Gavazzi \& Soucail 2007), and Suprime Cam, for which conservative estimates quote 35/arcmin$^2$ and higher (van Waerbeke et al. 2006). 

The pixel locations 
for each source are drawn from a uniform random distribution, while
each of the two components of the complex ellipticity, $\epsilon_i$, are
drawn from a random Gaussian distribution with a mean of zero and a
standard deviation of 0.2. The background sources are all placed at
the same redshift ($z_s=1.0$), and the brightness in the source plane generated according to
\begin{equation}
I(x,y)=\sum_i e^{-\frac{1}{2}\left(\frac{r_i(x,y)}{\sigma_i}\right)^{\frac{5}{2}}},
\end{equation}
where the sum is taken over all background galaxies,
\begin{eqnarray}
  r_i&=&\frac{\sqrt{\left((1-\epsilon_{1})x^\prime-\epsilon_{2}y^\prime\right)^2+
      \left(-\epsilon_{2}x^\prime+(1+\epsilon_1)y^\prime\right)^2}}
  {1-|{\mathbf\epsilon}|^2}, 
\end{eqnarray}
and $x^\prime=x-x_i$. We take $r_{\rm gal}=6$kpc to be a characteristic scale for the galaxies,  and define $\sigma=r_{\rm gal}/1.22d_{\rm gal}$, where $D_{s}$ is the angular diameter distance to the source plane.

The background galaxies are lensed by computing the mapping from each pixel location in the lens plane to the source plane and using a cubic interpolation to determine the brightness distribution at that location. 

\subsection{Computing the Aperture Mass Statistic}
To compute the aperture mass over the field of view, we use a grid of $1000\times1000$ apertures of a fixed radius. The flexion is measured using the weighted HOLICs formalism laid out in Goldberg \& Leonard (2007). As in Leonard et al. (2007), we reject any flexion measurements for which $a|\f|>0.2$, where $a$ is the measured semi-major axis of the lensed galaxy. This is done to ensure that no strongly lensed objects are included, and to reduce contamination from bad measurements or blended sources. 

In order to compute the noise in the aperture mass measurement, the data are randomised by rotating each flexion vector by an angle drawn randomly from a uniform distribution in the range $0\le \theta \le 2\pi$ and computing the aperture mass on the randomised data. This procedure is carried out 5000 times, and the noise is given by the standard deviation of the aperture mass maps generated using randomised data.

\section{Results}
\subsection{Single Lens Galaxy}
\label{subsec:simple}
As a first test of the aperture mass method, we consider here a single lens galaxy with a velocity dispersion of $500$km s$^{-1}$, corresponding to an Einstein radius of $\theta_E=5\arcsec .95$ assuming $D_{ls}/D_{s}=0.8$. In addition, we consider data in which the flexion is measured exactly; i.e. we randomly sample points on the lens plane grid at a rate of $35$ arcmin$^{-2}$, analytically compute the flexion at these points, and use these data to compute the aperture mass. Any data points that are found to lie within the Einstein radius of the lens are discarded prior to computing the aperture mass. 5000 randomisations of the data are carried out, the resulting dispersion giving a measurement of the noise associated with the aperture mass. We use the polynomial flexion filter function described in \S~\ref{subsubsec:98}, taking $l=3$ and $R=60\arcsec$. 

\begin{figure}
\center
\includegraphics[width=0.4\textwidth]{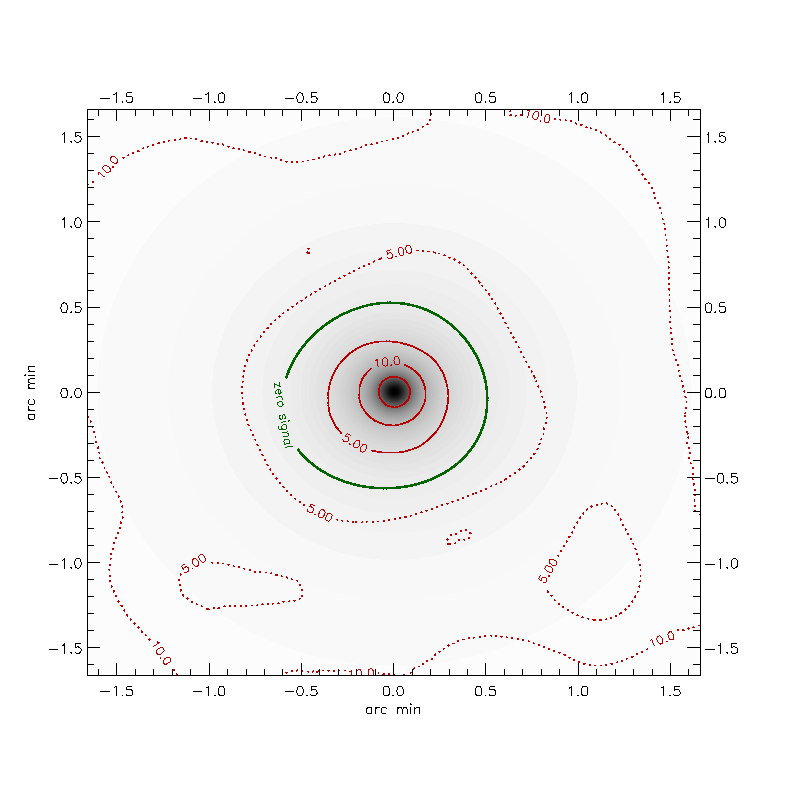}
\caption{The flexion aperture mass signal to noise contours for a simple softened isothermal lens and randomly-sampled theoretical data overlaid on a density plot of the underlying convergence. The solid contours show positive signal to noise, while the dashed contours show regions of negative signal. The zero point contour is clearly labelled. \label{fg:simple}} 
\end{figure}

Figure \ref{fg:simple} shows the signal to noise map for the aperture mass computed in this way. As expected from the calculation in \S~\ref{subsec:sn}, we find the peak signal to noise for this reconstruction to be ${\cal S}_{peak}=18.1$. A very noticeable feature of the aperture mass map generated in this way is that there is a clear radius at which the signal becomes negative. We expect the aperture mass to become negative at large radii as the mass filter function becomes negative, and the radius at which this occurs will be dependent on the mass profile of the lens, the strength of the lens, the shape of the filter function used, and the size of the aperture. This implies that if this zero contour can be resolved in real images, and a reasonable shape for the mass profile of the lens can be assumed, one might be able to infer the mass of the lens (or its Einstein radius) simply by measuring the radius of the aperture mass zero contour.

\begin{figure}
\center
\includegraphics[width=0.4\textwidth]{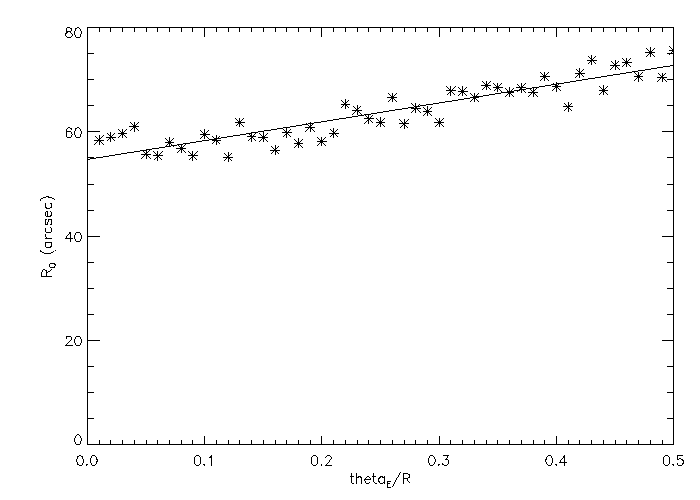}
\caption{The zero point radius $R_0$ as a function of $\theta_E/R$ for an aperture size of $R=120\arcsec$ and a polynomial order $l=5$. The discrete points represent the actual data, and the solid line shows the line of best fit for the data. The solid curves represent regions of positive signal, while the dashed curves show areas of negative signal. \label{fg:zero_pt}}
\end{figure}

A detailed treatment of this idea is beyond the scope of this paper; however, as a test of this concept, we measured the zero point radius of the aperture mass for a simple single softened isothermal lens of variable Einstein radius, and for various combinations of aperture scale and polynomial order. Figure \ref{fg:zero_pt} shows the zero point radius plotted as a function of $\theta_E/R$ for an aperture radius of $R=120\arcsec$ and a polynomial filter with $l=5$. It can clearly be seen from the figure that the relationship between $R_0$ and $\theta_E$ appears to be linear. Different combinations of $l$ and $R$ yield similar linear results, though with slightly different slopes as $l$ is varied. 

Clearly as the lens model becomes more complicated, or less smooth, the relationship between the Einstein radius and the zero point radius will become less straightforward. However, it does seem that if these zero point contours can be resolved using the flexion aperture mass, this might provide a straightforward way to put at the very least broad constraints on the mass contained within the structures seen.
\subsection{Analytic Simulation of a Group of Galaxies}

We now consider a group of softened isothermal profile galaxies generated as described in \S~\ref{subsec:analytic}. These galaxies were generated to lie within a region corresponding to a physical scale of $0.67$Mpc on a side at a redshift of $z_l=0.16$. The aperture mass statistic was calculated using both the polynomial and piecewise continuous filter functions described in \S~\ref{subsubsec:98} and \S~\ref{subsubsec:96}. In addition, our flexion data are generated using the method described in \S~\ref{subsec:bg}. 

\begin{figure}
\center
\includegraphics[width=0.4\textwidth]{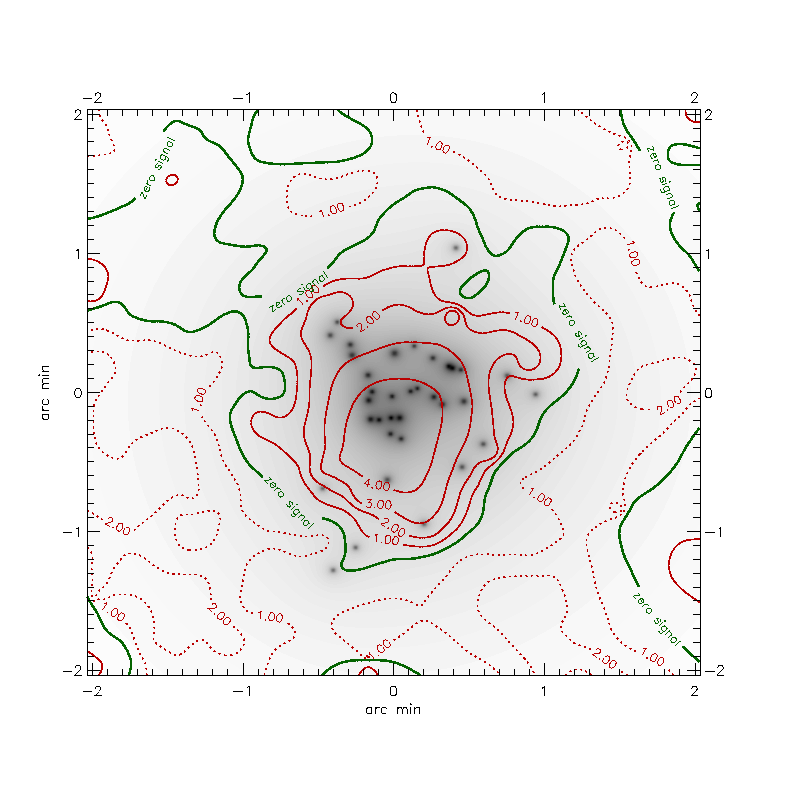}
\includegraphics[width=0.4\textwidth]{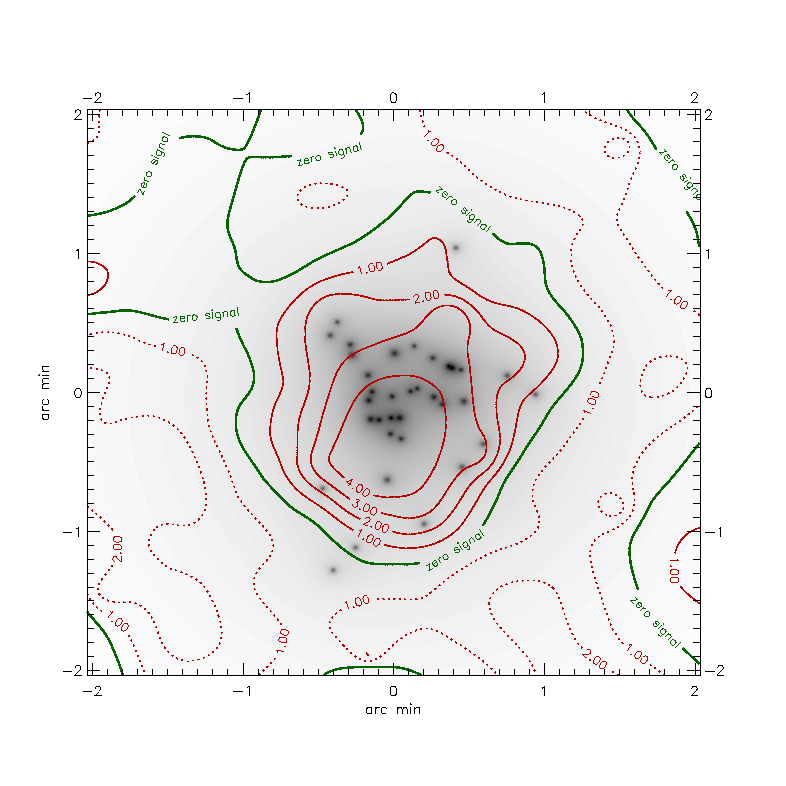}
\caption{The aperture mass signal to noise contours computed using simulated data and overlaid on an image of the underlying convergence. The aperture mass is computed using piecewise continuous filter functions with $\nu_1=0.05$ ({\textit Top Panel}) and $\nu_1=0.1$ ({\textit Bottom Panel}). For each, the outer radius is taken to be $\nu_2=0.95$. \label{fg:an_96}}
\end{figure}

Figure \ref{fg:an_96} shows the aperture mass signal to noise contours computed using the piecewise-continuous filter functions of \S~\ref{subsubsec:96} using an aperture scale of $R=90\arcsec$ and taking $\nu_1=0.05$ and $0.1$. With both values of the inner radius, the signal to noise contours are seen to quite accurately trace the shape of the convergence, to a peak signal to noise of ${\cal S}_{peak}=5.0$ for $\nu_1=0.05$, and ${\cal S}_{peak}=5.1$ for $\nu_1=0.1$. As expected, the signal to noise contours for the smaller inner radius resolve structures on smaller scales than that for larger inner radius. Indeed, reconstructions involving a different aperture size, though not shown here, exhibit the expected behaviour in terms of the scale on which structures are resolved. In addition, a zero point contour is quite clearly resolved in both signal to noise plots.  

\begin{figure}
\center
\includegraphics[width=0.4\textwidth]{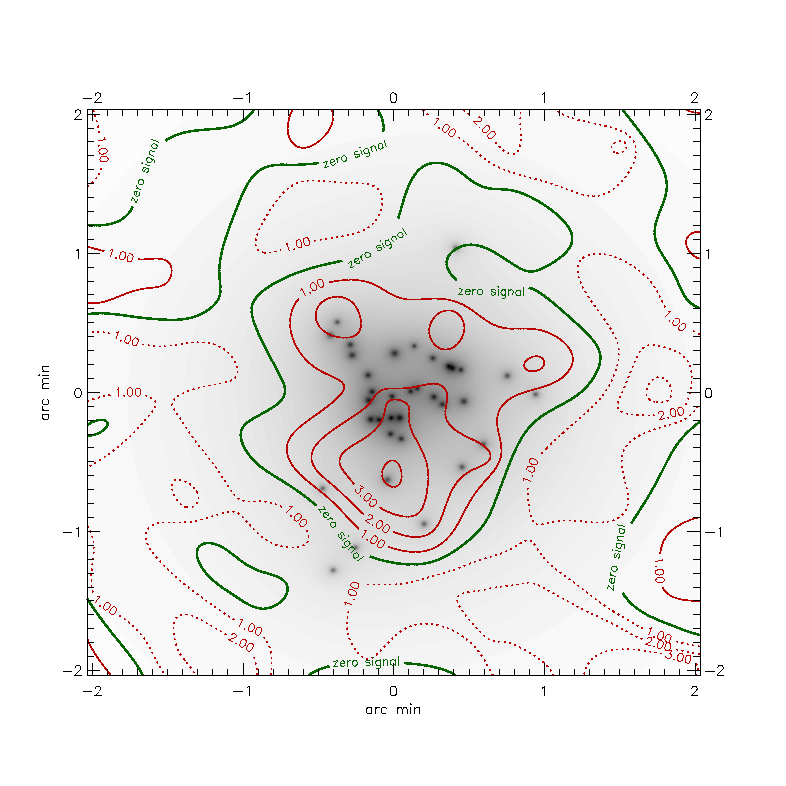}
\includegraphics[width=0.4\textwidth]{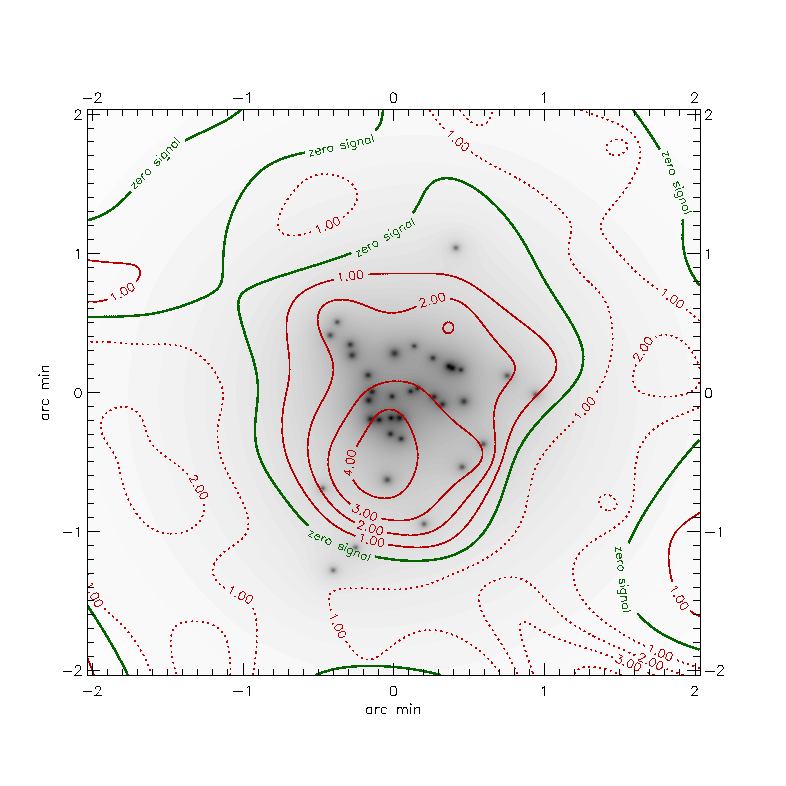}
\includegraphics[width=0.4\textwidth]{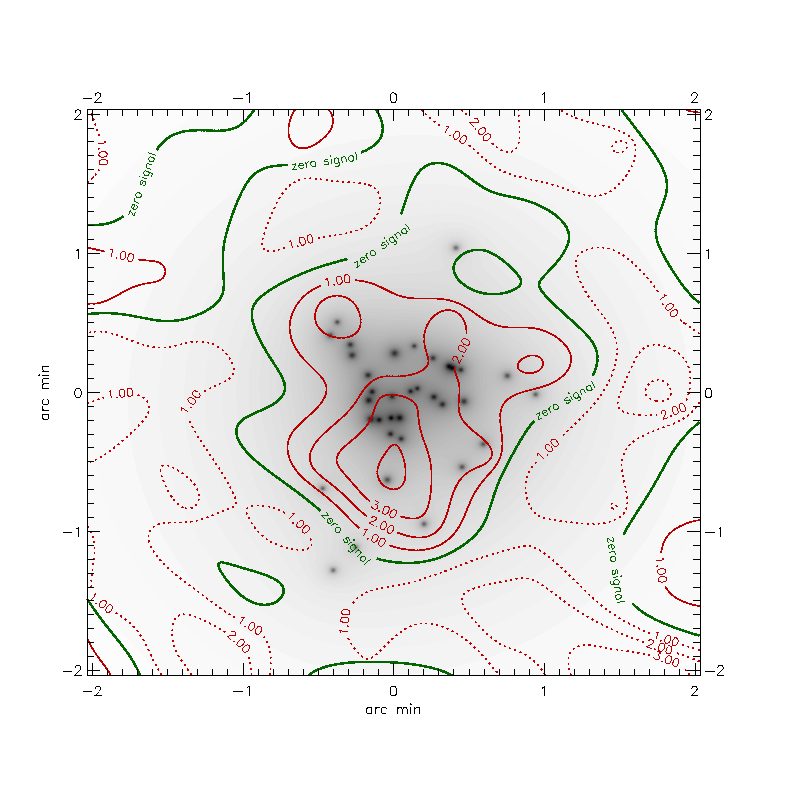}
\caption{The aperture mass signal to noise contours computed using simulated data and overlaid on an image of the underlying convergence. The aperture mass is computed using polynomial filter functions with $l=3,\ R=60\arcsec$ ({\textit Top Panel}), $l=5,\ R=90\arcsec$ ({\textit Middle Panel}), and $l=10\ R=90\arcsec$ ({\textit Bottom Panel}). The solid curves represent regions of positive signal, while the dashed curves represent regions of negative signal. \label{fg:an_98}}
\end{figure}
Figure \ref{fg:an_98} shows results obtained using the polynomial filter functions, shown for the combinations $[R,l]=[60\arcsec,3],[90\arcsec,5]$ and $[90\arcsec,10]$. The peak signal to noise obtained in these reconstructions is ${\cal S}_{peak}=4.1,\ 4.5$ and $4.2$, respectively. Here again the signal to noise contours accurately trace the shape of the galaxy group, and are able to resolve structures on subarcminute scales. In addition, as expected, we see that increasing the aperture size increases the minimum scale on which substructure can be resolved, and increasing the polynomial order allows us to probe these smaller-scale substructures. Further, we again resolve a zero-point contour, the radius of which, when compared to the linear relation shown in \S~\ref{subsec:simple}, yields an Einstein radius which is broadly consistent (within a factor of $\sim 2-3$) with that found by modelling all the mass contained in the lens as a single softened isothermal sphere.

\begin{figure}
\center
\includegraphics[width=0.4\textwidth]{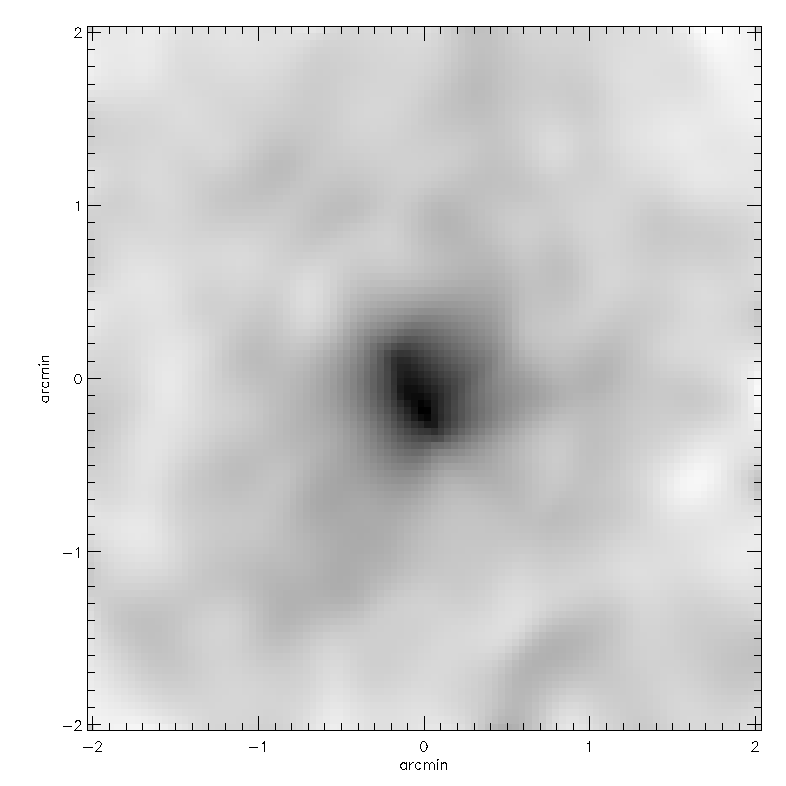}
\caption{A reconstruction of the convergence from shear data using the finite field inversion technique of Seitz \& Schneider (2001); see text for details.\label{fg:shear}}
\end{figure}
As a check on the performance of this method, we compare our results to a weak lensing mass reconstruction, by using the measured ellipticities of our lensed images to perform a finite field inversion following the method of Seitz and Schneider (2001). The reconstruction was performed on a $100\times 100$ grid, with a smoothing scale of 120 pixels (0.048 times the image size) - chosen to be large enough to include averaging over sufficient background galaxies to lessen the noise from their intrinsic 
ellipticity dispersion. The resulting mass reconstruction is shown in figure \ref{fg:shear}. This reconstruction is clearly unable to resolve the small-scale structure of this galaxy group, or to accurately trace the overall shape of the convergence distribution.

\subsection{Cluster of Galaxies from N-body Simulations}

\begin{figure}
\center
\includegraphics[width=0.4\textwidth]{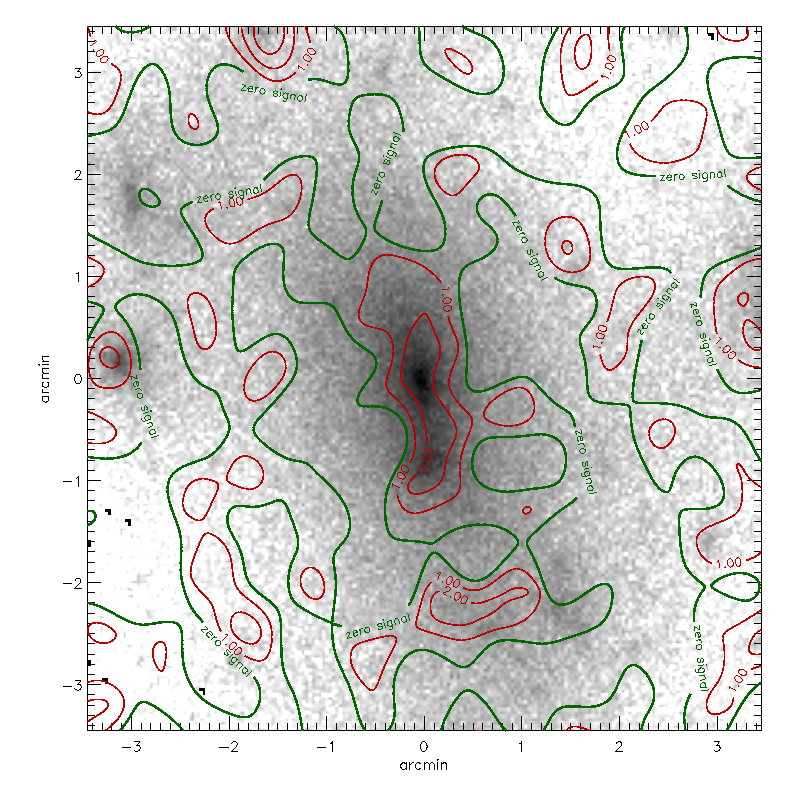}
\includegraphics[width=0.4\textwidth]{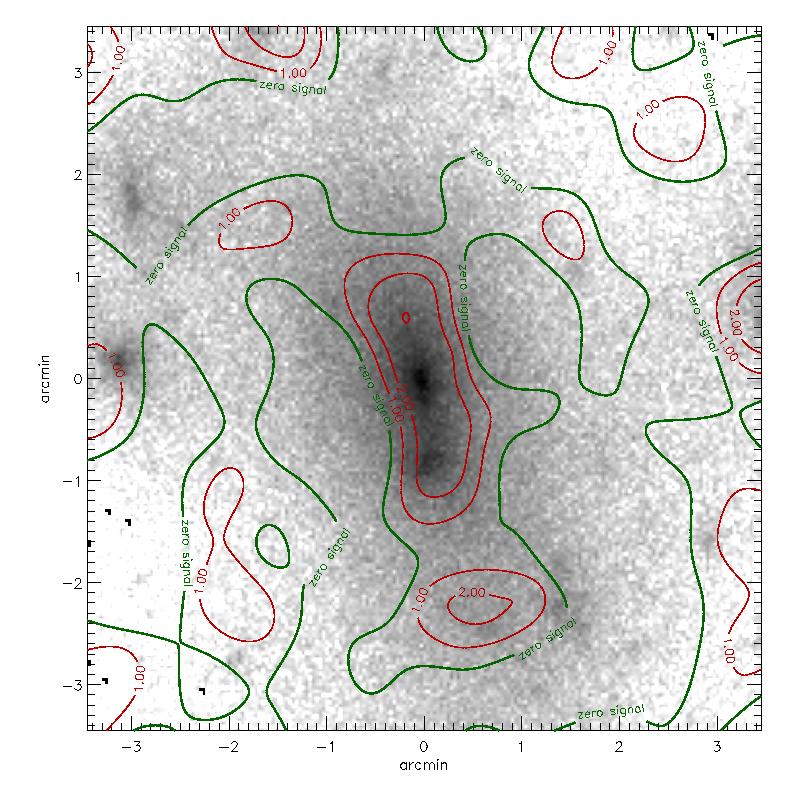}
\includegraphics[width=0.4\textwidth]{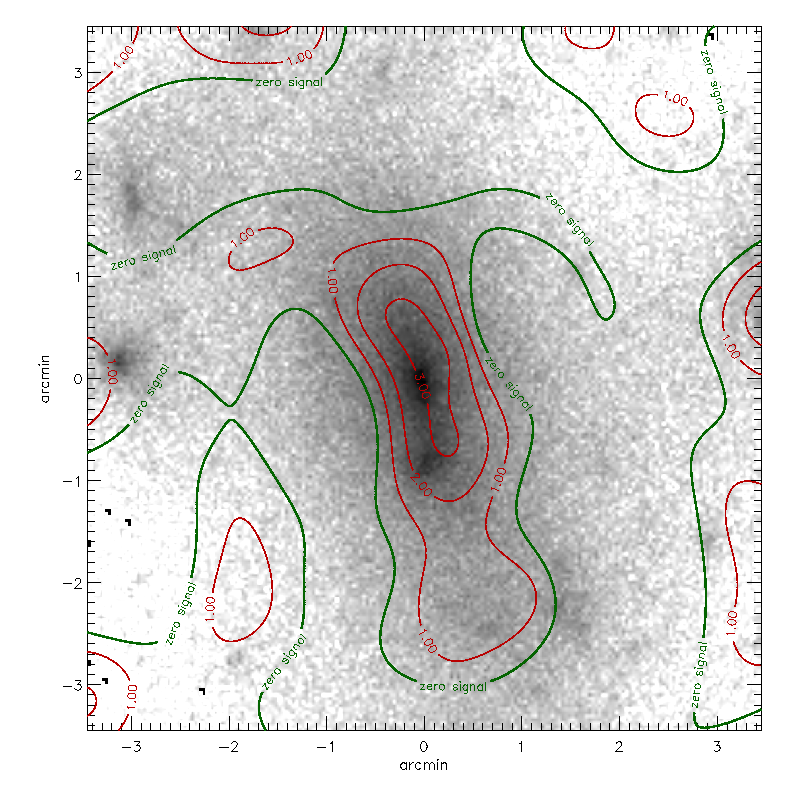}
\caption{Aperture mass signal to noise contours computed using simulated data lensed through a cluster from the Millennium simulation, overlaid on the underlying convergence. Only positive signal to noise contours are plotted. The aperture mass was calculated using a polynomial filter function with $l=5$, and with aperture radii of $R=60\arcsec$ (\textit{top panel}), $R=90\arcsec$ (\textit{middle panel}), and $R=120\arcsec$ (\textit{bottom panel}). \label{fg:nbody}}
\end{figure}

Finally, we consider a cluster of galaxies taken from the Millennium simulation and described in \S~\ref{subsec:nbody}. Figure \ref{fg:nbody} shows the signal to noise contours for the flexion aperture mass statistic applied to this cluster using the polynomial filter functions with $l=5$ and $R=60\arcsec,\ 90\arcsec$ and $120\arcsec$. The peak signal to noise for these maps is found to be ${\cal S}_{peak}=3.8,\ 3.0,$ and $3.3$, respectively. Similar results were obtained using the piecewise-continuous filter functions.

We expect, and indeed find, the signal to noise map to be somewhat noisier than that seen with the analytic simulations. This manifests as spurious detections on small scales of structures that do not exist, which undoubtedly result from using rather noisy data to compute the deflection angles. As expected, when the aperture scale is increased, these detections are removed, as the aperture mass becomes less sensitive to structures on those scales. 

While considerably noisier, the flexion aperture mass does still accurately trace the shape of the central part of the cluster, and resolves two smaller substructures (one to the left and one directly below the centre of the cluster) in all three signal to noise maps. However, as found in Leonard et al. (2007), the flexion measurements are generally quite insensitive to the smooth component of the cluster potential, so outside of the central, cuspy region, the flexion signal drops off rather quickly. 

\begin{figure}
\center
\includegraphics[width=0.4\textwidth]{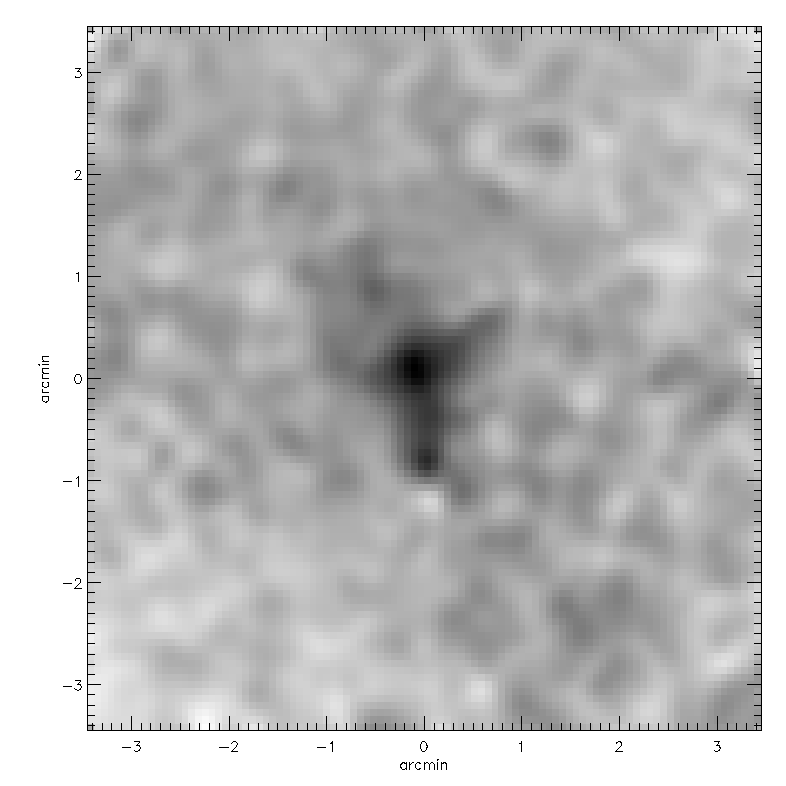}
\caption{Shear reconstruction of the convergence field using the finite field inversion technique described in Seitz and Schneider (2001); see text for details. \label{fg:nbodyshear}}
\end{figure}

Again, we can compare the results from flexion with that from shear. As for the analytic group, we use the Seitz and Schneider (2001) finite field inversion technique to reconstruct the convergence from our shear data. The mass was reconstructed on a $100\times 100$ grid, with a smoothing scale of 0.029 times the image size. The mass reconstruction is shown in figure \ref{fg:nbodyshear}. In this case, the shear does pick up the overall shape of the cluster quite accurately, but does not clearly resolve the smaller substructures seen in the flexion aperture mass signal to noise map.

\section{Discussion and Conclusions}

In this paper, we have derived an aperture mass statistic for (first) flexion, and shown that it provides a robust method for detecting structures within clusters of galaxies on sub-arcminute scales. Whilst the filter functions used here are by no means optimal, they offer an excellent proof of concept for this technique, and show that even using non-optimal filters, and a moderate number density of background galaxies, mass concentrations can be detected on a range of different scales with an appreciable peak signal to noise that varies very little with changes to the aperture scale radius or filter parameters ($\nu_1$, $\nu_2$, and $l$). 

We have also shown that the zero point contours of the aperture mass map can be clearly resolved in simulated data, and described briefly how these zero points might be used to provide an estimate of the mass contained in structures identified using this method. This is, of course, model- and filter-dependent, but might provide a good starting point for maximum-likelihood modelling in clusters of galaxies where flexion constraints are incorporated to describe the substructure within the cluster, for example. 

It is clear that the filters used in the preceding analysis are not optimised for any particular lens, but can be used to detect structures with a range of different mass profiles. This will be useful for detecting substructures where one has little or no prior knowledge of the mass profiles. However, it is clear that improvements can be made to the method, and improved signal to noise obtained, by using filters that are tailored to the structures being studied. This will be particularly important when dealing with noisier data, or data for which the background source number density is low, thus reducing the maximum attainable signal to noise.

Our understanding of the physical properties of galaxy clusters and their utility as cosmological probes will be greatly enhanced with the advent of large surveys such as the Dark Energy Survey (DES), augmented with complementary data from experiments such as the South Pole Telescope (SPT).
For the large cluster samples, gravitational flexion will provide constraints on the mass substructure of clusters at a higher resolution than weak lensing, and probes greater distances from the cluster centre than strong lensing. Clearly, a priority is to simultaneously use lensing data from various regimes in conjunction with other types of data to obtain a detailed multi-scale view of the largest bound structures in the universe.

\section{acknowledgments}
We thank Anthony Challinor, Antony Lewis, Gordon Ogilvie and Peter Schneider for discussions. 
We thank John Helly and Ian McCarthy for providing the Millennium Simulation cluster particle data.
AL is supported by a BP/STFC Dorothy Hodgkin Postgraduate Award, LJK is supported by a Royal Society University Research Fellowship, and SMW is supported by an STFC Postgraduate award.

\appendix

\section[]{Relationship Between the Convergence and Flexion}

\label{ap:kernels}
The convergence can be related to the measured flexion in Fourier
space by (Bacon et al. 2006):
\begin{eqnarray}
\tilde{\kappa} &=& \frac{ik_1}{k_1^2+k_2^2}\tilde{{\cal
 F}_1}+\frac{ik_2}{k_1^2+k_2^2}\tilde{{\cal F}_2}\nonumber\\
&=& \tilde{E}_1^\f\tilde{\f}_1+\tilde{E}_2^\f\tilde{\f}_2.
\label{eq:Fkern}
\end{eqnarray}

This implies that the real-space relationship between $\kappa$ and the
flexion is a convolution:
\begin{equation}
\kappa=\frac{1}{2\pi}\left(E_1^{\f\ast}\otimes \f_1+E_2^{\f\ast}\otimes \f_2\right),
\end{equation}
or
\begin{equation}
\kappa(\bx)=\frac{1}{2\pi}\Re\left[\int d^2\bx^\prime
  \f(\bx^\prime)E^\ast_\f(\bx-\bx^\prime)\right],
\end{equation}
where $E_\f$ is a convolution kernel to be determined.

Noting that 
\begin{equation}
F\left(\frac{ik_j}{|{\bf k}|^2}\right)=\frac{x_j}{2\pi|\bx|^2},
\end{equation}
where $F(f({\bf k}))$ denotes the fourier transform of the function
$f({\bf k})$, we find that
\begin{equation}
E_\f(X)=\frac{1}{X^\ast},
\end{equation}
where $X=x_1+ix_2=xe^{i\phi}$, and $\phi$ is the phase of the complex
number $X$.

\section[]{Expressing the Aperture mass statistic in terms of the E-mode flexion}
\label{ap:filt}

The aperture mass is defined, in terms of the measured flexion, as 
\begin{eqnarray}
\lefteqn{m(\bx_0)=}\nonumber\\
& &\Re\left[\int d^2\bx^\prime \f(\bx^\prime)\int d^2\bx\ 
  E_\f^\ast(\bx+\bx_0-\bx^\prime)w(|\bx|)\right]. 
\end{eqnarray}
Making the transformation $\by=\bx^\prime-\bx_0$, and writing $d^2\bx
= x\ dx\ d\phi$, we obtain
\begin{eqnarray}
\label{eq:starthere}
\lefteqn{m(\bx_0)=}\nonumber\\
& &\Re\left[\int d^2\by\ \f(\by+\bx_0)\int_0^\infty x\ w(x)dx
  \int_0^{2\pi}d\phi\ E_\f^\ast(X-Y)\right],\nonumber\\
\end{eqnarray}
where $X=x_1+ix_2=|\bx|e^{i\phi}$. 
We consider the rightmost integral first, which can be rewritten as 
\begin{equation}
I_1=\int_0^{2\pi}\frac{d\phi}{2\pi(X-Y)}.
\end{equation} 
This integral has a regular singularity at $X=Y$, and using the
transformation $dX=iXd\phi$, can be expressed as a contour integral:
\begin{equation}
I_{1}=-\frac{i}{2\pi}\oint \frac{dX}{X(X-Y)},
\end{equation}
where the integration is carried out along a circular contour of
radius $|X|$. This integrand has two simple poles at $X=0$ and
$X=Y$. The residues at these singularities are given by
\begin{equation}
Res(X=0)= -\frac{1}{Y},\end{equation}\begin{equation}
Res(X=Y)= \ \frac{1}{Y}.
\end{equation}
Thus, applying the residue theorem, 
\begin{equation}
I_{1}=-\frac{1}{Y}\Theta(|Y|-|X|),
\end{equation}
where
\begin{equation}
\Theta(x)=\begin{cases} 1 & x>0\\ 0 & x<0\end{cases}.
\end{equation}
Returning to equation \ref{eq:starthere}, we find that
\begin{eqnarray}
m(\bx_0)&=&-\Re\left[\int d^2\by\  \f(\by+\bx_0)\frac{1}{Y}\int_0^\infty x\ w(x)
  \Theta(y-x)\ dx\right]\nonumber\\ 
&=&-\Re\left[\int d^2\by\  \f(\by-\bx_0)\frac{1}{Y}\int_0^y x\ w(x)\ dx
\right].\nonumber\\ 
\end{eqnarray}

Now, the E-mode flexion is given by
\begin{eqnarray}
\f_E&=&\Re\left[\f e^{-i\phi}\right]\nonumber\\
&=& \Re\left[\f \frac{|Y|}{Y}\right],
\end{eqnarray}
thus we can write
\begin{equation}
m(\bx_0)=-\int d^2\by\  \f_E(\by-\bx_0)\left[\frac{1}{y}\int_0^y\ x\ w(x)\ dx\right],
\end{equation}
which leads directly to equations \ref{eq:mflex} and \ref{eq:qwflex}.

Thus, we have expressed $Q_\f(x)$ in terms of $w_\f(x)$ as 
\begin{equation}
Q_\f(x)=-\frac{1}{x}\int_0^x\ y\ w(y)\ dy.
\end{equation}
We now aim to derive the reverse relation; i.e. to express $w_\f(x)$
as a function of $Q_\f(x)$.

Recall that first flexion is related to the convergence via
\begin{equation}
\f=\f_1+i\f_2=\partial\kappa=(\partial_1+i\partial_2)\kappa.
\end{equation}
It is useful to transform the derivative operators into polar
coordinates as
\begin{eqnarray}
\partial_1=\cos\phi\partial_r-\frac{\sin\phi}{r}\partial_\phi,\nonumber\\
\partial_2=\sin\phi\partial_r+\frac{\cos{\phi}}{r}\partial_\phi.
\end{eqnarray}
Using the above expressions, and recalling that the E-mode flexion is
defined as $\f_E\equiv\f_1\cos\phi+\f_2\sin\phi$, we find that the
E-mode flexion is related quite simply to the convergence via
$\f_E=\partial_r\kappa$. Thus, we can express the aperture mass
statistic as
\begin{eqnarray}
m(\bx_0)&=&\int d^2\bx\left(\frac{\partial\kappa}{\partial
    r}\right)_{\bx+\bx_0} Q_\f(y)\nonumber\\
&=& \int_0^{2\pi}d\phi\int_0^\infty r\frac{\partial\kappa}{\partial r}\ dr.
\end{eqnarray}
Integrating the inner expression by parts and simplifying, we find that 
\begin{equation}
  m(\bx_0)=-\int d^2\bx\ 
  \kappa(\bx+\bx_0)\left(\frac{1}{r}Q_\f(r)+\frac{dQ_\f}{dr}\right). 
\end{equation}
Comparing this expression to equation \ref{eq:apmass}, it follows that
\begin{equation}
w_\f(r)=-\frac{1}{r}Q_\f(r)-\frac{dQ_\f}{dr}.
\end{equation}

%
\label{lastpage}

\end{document}